\documentclass{llncs} % instead of \documentclass[11pt]{article}
\newtheorem{defn}{Definition} %[section]
\newtheorem{df}{Definition}%[defn]

\newtheorem{thm}{Theorem}
\newtheorem{lem}[defn]{Lemma}
\newtheorem{cor}{Corollary}
\newtheorem{rem}{Remark}
\newtheorem{con}{Conjecture}

\title{The \v{C}erny conjecture}

\date{11.07.2018}
\author{A.N. Trahtman\thanks{Email: avraham.trakhtman@gmail.com}
\institute{}
}

\begin{document}
\maketitle
\begin{abstract}
  A word $w$ of letters on edges of underlying graph $\Gamma$ of deterministic finite automaton (DFA)
is called synchronizing if $w$ sends all states of the automaton to a unique state.
 J. \v{C}erny discovered in 1964 a sequence of $n$-state complete DFA
possessing a minimal synchronizing word of length $(n-1)^2$.

The hypothesis, well known today as the \v{C}erny conjecture, claims that it is also precise
 upper bound on the length of such a word for a complete DFA. The hypothesis was formulated in 1966 by Starke.
The problem has motivated great and constantly growing number of investigations and generalizations.

To prove the conjecture, we use algebra w on a special class of row monomial matrices
(one unit and rest zeros in every row), induced by words in the alphabet of labels on edges.
These matrices generate a space with respect to the mentioned operation.

The proof is based on connection between length of words $u$
and dimension of the space generated by solutions $L_x$
of matrix equation $M_uL_x=M_s$ for synchronizing word $s$,
as well as on the relation between ranks of $M_u$ and $L_x$.

\end{abstract}

$\bf Keywords$: deterministic finite automaton, synchronizing word, \v{C}erny conjecture.
\section*{Introduction}

 The problem of synchronization of DFA is a natural one and various aspects of this problem
have been touched in the literature.
Prehistory of the topic, the emergence of the term, the connections with the early coding theory,
first efforts to estimate the length of synchronizing word \cite{La}, \cite{Li},
different problems of synchronization one can find in surveys \cite{Ju}, \cite{KV}.

Synchronization makes the behavior of an automaton resistant
against input errors since, after detection of an error,
a synchronizing word can reset the automaton back to its original
state, as if no error had occurred.
The synchronizing word limits the propagation of errors for a prefix code.
Deterministic finite automaton is a tool that helps to recognized language in a set of DNA strings.

A problem with a long story is the estimation of the minimal length of synchronizing word.

 J. \v{C}erny in 1964 \cite{Ce} found the sequence of $n$-state complete DFA with shortest
synchronizing word of length $(n-1)^2$ for an alphabet of size two.
  The hypothesis, well known today as the \v{C}erny's conjecture, claims that this lower bound
on the length of the synchronizing word of aforementioned automaton is also the upper bound
for the shortest synchronizing word of any $n$-state complete DFA:
\begin{con}
The deterministic complete $n$-state synchronizing automaton over alphabet $\Sigma$
has synchronizing word in $\Sigma$ of length at most $(n-1)^2$ \cite{Sta} (Starke, 1966).
  \end{con}

The problem can be reduced to automata with a strongly connected graph \cite{Ce}.
An attempt to prove this hypothesis is proposed below.

This famous conjecture is true for a lot of automata, but in general the problem still remains
open although several hundreds of articles consider this problem from different points of view \cite{TB}.

Moreover, two conferences "Workshop on Synchronizing  Automata"
(Turku, 2004) and  "Around the  \v{C}erny conjecture"
(Wroclaw, 2008) were dedicated to this longstanding conjecture.
The problem is discussed in "Wikipedia" - the popular Internet
Encyclopedia and on many other sites.

As well as the Road Coloring problem \cite{AW}, \cite{Fi}, \cite{TP}, this simple-looking
conjecture was arguably  the most longstanding and famous open combinatorial problems
in the theory of finite automata \cite{KV}, \cite{MS}, \cite{PS}, \cite{Sta}, \cite{St}.

We consider a class of matrices $M_u$ of mapping induced by words $u$ in the alphabet
of letters on edges of the underlying graph $\Gamma$. The matrix $M_u$ of word $u$
 belongs to the class of matrices with one unit in every row and rest zeros  (row monomial).
We call them also matrices of word.

There are no examples of automata such that the length of
the shortest synchronizing word is greater than $(n-1)^2$.
Moreover, the examples of automata  with shortest synchronizing
word of length $(n-1)^2$ are infrequent.
After the sequence of \v{C}erny and the example of \v{C}erny, Piricka and Rosenauerova \cite{CPR} of 1971
for $|\Sigma|=2$, the next such examples were found by Kari \cite {Ka} in 2001 for $n=6$ and $|\Sigma|=2$
and by Roman \cite {Ro} for $n=5$ and $|\Sigma|=3$ in 2004.

The package TESTAS \cite {TP}, \cite {Tt} studied all automata with strongly connected underlying graph
of size $n \le 11$ for  $|\Sigma|=2$, of size $n \le 8$ for $|\Sigma| \le 3$ and of size $n \le 7$ for $|\Sigma| \le 4$
and found five new examples of DFA with shortest synchronizing word of length $(n-1)^2$ with $n\leq 4$.
 Don and Zantema present in \cite {DZ} an ingenious method of designing new automata from existing examples
of size three and four and proved that for $n\geq 5$ the method does not work.
So there are up to isomorphism exactly 15 DFA for $n=3$ and exactly 12 DFA for $n=4$
 with shortest synchronizing word of length $(n-1)^2$.

The authors of \cite {DZ} support the hypothesis from \cite{TS} that all automata with shortest synchronizing word
of length $(n-1)^2$ are known, of course, with essential correction found by  themselves for $n=3,4$.

There are several reasons \cite{AGV}, \cite{BBP}, \cite{CA},  \cite {DZ}, \cite{TS} to believe that the length
of the shortest synchronizing word for remaining automata  with $n>4$ (except the sequence of \v{C}erny
and examples for $n=5, 6$) is essentially less and the gap grows with $n$.
For several classes of automata, one can find some estimations on the length in \cite{AGV}, \cite{CR}, \cite{Kr},
\cite{KKS}, \cite{Ta}.

Initially found upper bound for the minimal length of synchronizing word was big and
has been consistently improved over the years by different authors.
The upper bound found by Frankl in 1982 \cite{Fr} is equal to $(n^3-n)/6$.
The result was reformulated in terms of synchronization in \cite{Pin} and repeated independently in \cite{KRS}.
The cubic estimation of the bound exists since 1982. Attempts to improve Frankl's result were unsuccessful.

The considered deterministic automaton $A$ can be presented by a
complete underlying graph with edges labelled by letters of an alphabet.

Our work uses the class of row monomial matrices $M_u$ of mapping induced by words $u$ in the alphabet
of letters on edges of the underlying graph and properties of corresponding space.

The matrix approach for synchronizing  automata supposed first by B{\'e}al \cite{Be} proved to be fruitful
\cite{BBP}, \cite{CA}, \cite{Co}.

There are no examples of automata such that the length of the shortest synchronizing word is greater than $(n-1)^2$.
Moreover, the examples of automata  with shortest synchronizing word of length $(n-1)^2$ are infrequent.
After the sequence of \v{C}erny and the example of \v{C}erny, Piricka and Rosenauerova \cite{CPR} of 1971
for $|\Sigma|=2$, the next such examples were found by Kari \cite {Ka} in 2001 for $n=6$ and $|\Sigma|=2$
and by Roman \cite {Ro} for $n=5$ and $|\Sigma|=3$ in 2004.

There are several reasons \cite{AGV}, \cite{BBP}, \cite{CA},  \cite {DZ}, \cite{TS} to believe that the length
of the shortest synchronizing word  for remaining automata  with $n>4$ (except the sequence of \v{C}erny
and two examples for $n=5, 6$) is essentially less and the gap grows with $n$.

We consider the equation $M_uL_x=M_s$ (\ref{ux}) for synchronizing word  $s$ and the space generated
by row monomial solutions $L_x$.
A connection between the set of nonzero columns of matrix of word, subsets of states of automaton
and our kind  $L_x$ of solutions  of (\ref{ux}) is revealed in Remarks.

Theorems  \ref{t}, \ref{t2} finish our attempt to prove the \v{C}erny conjecture.
Theorem \ref{t4} and some corollaries contain certain consequences.
The ideas of the proof are illustrated on example of automata with a maximal length of synchronizing word from \cite{Ka}.

 \section*{Preliminaries}
We consider a complete $n$-state DFA with
 strongly connected underlying graph $\Gamma$ and transition semigroup $S$
 over a fixed finite alphabet $\Sigma$ of labels on edges of  $\Gamma$ of an automaton $A$.
The trivial cases $n \leq 2$, $|\Sigma|=1$ and $|A \sigma|=1$ for
$\sigma \in\Sigma$ are excluded.

The restriction on strongly connected graphs is based on \cite{Ce}.
The states of the automaton $A$ are considered also as vertices of the graph $\Gamma$.

If there exists a path in an automaton from the state $\bf p$ to
the state $\bf q$ and the edges of the path are consecutively
labelled by $\sigma_1, ..., \sigma_k$, then for
$s=\sigma_1...\sigma_k \in \Sigma^+$ let us write ${\bf q}={\bf p}s$.

Let $Px$ be the set of states ${\bf q}={\bf p}x$ for all ${\bf p}$
from the subset $P$ of states and $x \in \Sigma^+$.
Let $Ax$ denote the set $Px$ for the set $P$ of all states of the automaton.

 A word $s \in \Sigma^+ $ is called a {\it synchronizing (reset, magic, recurrent, homing, directable)}
 word of an automaton $A$ with underlying graph $\Gamma$ if $|As|=1$.
The word $s$ below denotes minimal synchronizing word such that for a state $\bf q$ $As=\bf q$.

The states of the automaton are enumerated, the state $\bf q$  has number one.

 An automaton (and its underlying graph) possessing a synchronizing word is called {\it synchronizing}.

Let us consider a linear space generated by row monomial (one unit and rest of zeros in every row)
$n\times n$-matrices.

We connect a mapping of the set of states of the automaton made by
a word $u$ with an $n\times n$-matrix $M_u$ such that for an element $m_{i,j} \in M_u$ takes place

\centerline{$m_{i,j}$= $\cases{1, &${\bf q}_i u ={\bf q}_j$; \cr 0, &otherwise.}$}

Any mapping of the set of states of the automaton  $A$  can be presented by some row monomial word $u$
and by a corresponding matrix $M_u$.
For instance,

 \centerline{$M_u=\left(
\begin{array}{ccccccc}
  0 & 0 & 1 & . & . & . &  0 \\
  1 & 0 & 0 & . & . & . &  0 \\
  0 & 0 & 0 & . & . & . &  1 \\
  . & . & . & . & . & . &  . \\
  0 & 1 & 0 & . & . & . &  0 \\
  1 & 0 & 0 & . & . & . &  0 \\
\end{array}\right)
$}

 Let us call the matrix $M_u$ of the mapping induced by the word $u$, for brevity, the matrix of word $u$.

$M_uM_v=M_{uv}$ \cite{Be}.

The set of nonzero columns of $M_u$ (set of second indexes of its elements) of $M_u$ is denoted as $R(u)$.

Zero matrix is considered as a matrix of empty word.

The subset of states $Au$ is denoted as $c_u$ with number of states $|c_u|$.
In $n$-vector $c_u$ the coordinate $j$ has unit if the state $j \in c_u$ and zero in opposite case.

For linear algebra terminology and definitions, see \cite{Ln}, \cite{Ma}.

\section{Mappings induced by a word and subword}

\begin{rem} \label{r1}
The invertible matrix $M_a$ does not change the number of units of every column of $M_u$ in its image of
the product $M_aM_u$.

Every unit in the product $M_uM_a$ is the product of two units, first unit from nonzero column of $M_u$
and second unit from a row with one unit of $M_a$.

\end{rem}

\begin{rem} \label{r4}

The columns of the matrix $M_uM_a$ are obtained by permutation of columns $M_u$.
Some columns can be merged (units of columns are moved along
row to a common column) with $|R(ua)|<|R(u)|$.

The rows of the matrix $M_aM_u$ are obtained by permutation of rows of the matrix $M_u$.
Some of these rows may disappear and replaced by another rows of $M_u$.

\end{rem}

\begin{lem} \label{l1}

The number of nonzero columns $|R(b)|$ is equal to the rank of $M_b$.

\centerline{$|R(ua)| \leq |R(u)|$ and   $R(au) \subseteq R(u)$.}

For invertible matrix $M_a$ $R(au)=R(u)$ and $|R(ua)|=|R(u)|$.

For the set of states of deterministic finite automaton $A$ and any words $u$ and $a$ $Aua \subseteq Aa$.

 Nonzero columns of $M_{ua}$ have units also in $M_a$.

\end{lem}

\begin{proof}
The matrix $M_b$ has submatrix with nonzero determinant having only one unit in every row and in every nonzero column
.
Therefore $|R(b)|$ is equal to the rank of $M_b$.

The matrix $M_a$ in the product $M_uM_a$ shifts column of
$M_u$ to columns of $M_uM_a$ without
changing the column itself by Remark \ref{r4} or merging.
some columns of $M_u$.

In view of possible merged columns, $|R(ua)|\leq |R(u)|$.

Some rows of $M_u$ can be replaced in $M_aM_u$ by another row and therefore some rows from $M_u$ may be changed,
but zero columns of $M_u$ remain in $M_aM_u$ (Remark 1).

Hence $R(au) \subseteq R(u)$ and $|R(ua)| \leq |R(u)|$.

For invertible matrix $M_a$ in view of existence $M_a^{-1}$ we have $|R(ua)|=|R(u)|$ and $R(au)= R(u)$.

From $R(ua) \subseteq R(a)$ follows $Aua \subseteq Aa$.

Nonzero columns of $M_{ua}$ have units also in $M_a$ in view of $R(ua) \subseteq R(a)$.
\end{proof}

\begin{cor}  \label{c1a}
The invertible matrix $M_a$ keeps the number of units of any column of $M_u$ in corresponding column of the product
$M_aM_u$.
\end{cor}

\begin{cor}  \label{c1}
The matrix $M_s$ of word $s$ is synchronizing if and only
if $M_s$ has zeros in all columns except one and units in
the residuary column.

All matrices of right subwords of $s$ also have at least one unit in this column.
\end{cor}

\section{Necessary conditions of the operation of summation in the class of row monomial matrices}

\begin{lem}\label{lam}
 Suppose that for row monomial matrices $M_i$
 and $M$
\begin{equation}
M =\sum_{i=1}^k\lambda_i M_i. \label{lm}
\end{equation}
with coefficients $\lambda$ from $Q$.

Then the sum $\sum^k_{i=1}\lambda_i =1$ and the sum $S_j$ of values in every row $j$
of the sum in (\ref{lm}) also is equal to one.

If $\sum^k_{i=1}\lambda_iM_i=0$  then $\sum_{i=1}^k \lambda_i=0$ and $S_j=0$
for every $j$ with $M_u=0$.

If the sum $\sum^k_{i=1}\lambda_i$ in every row is not unit
[zero] then $\sum_{i=1}^k\lambda_i M_i$
is not a row monomial matrix.
\end{lem}

\begin{proof}
The nonzero matrices $M_i$ have $n$ cells with unit in the cell.
Therefore, the sum of values in all cells of the matrix $\lambda_i M_i$ is $n \lambda_i$.

For nonzero $M$ the sum is $n$. So one has in view of
$M =\sum_{i=1}^k\lambda_i M_i$

\centerline {$n=n\sum_{i=1}^k \lambda_i$, whence $1 =\sum_{i=1}^k \lambda_i$.}
Let us consider the row $j$ of matrix $M_j$ in (\ref{lm}) and let  $1_j$ be unit in the row $j$.
The sum of values in a row of the sum (\ref{lm}) is equal to unit in the row of $M$.
So $1 =\sum_{i=1}^k \lambda_i1_i=\sum_{i=1}^k \lambda_i$.

$\sum_{i=1}^k\lambda_i M_i=0$ implies $S_j=\sum_{i=1}^k \lambda_i1_i=\sum_{i=1}^k \lambda_i=0$
for  every row $j$.

If the matrix $M=\sum_{i=1}^k\lambda_i M_i$ is a matrix
of word or zero matrix then
$\sum^k_{i=1}\lambda_i \in \{0, 1\}$.
If $\sum^k_{i=1}\lambda_i\not\in \{0, 1\}$ or
the sum  in ${0, 1}$ is not the same in every row then we have opposite case and
the matrix does not belong  to the set of row monomial matrix.
\end{proof}

The set of row monomial matrices is closed with respect
to the considered operation and together with zero matrix generates a space.

\section{Useful lemmas}

\begin{lem}  \label {v3}
 The set $V$ of all $n\times k$-matrices of words
(or $n\times n$-matrices with zeros in fixed $n-k$ columns for $k<n$) has $n(k-1)+1$ linear independent matrices.
 \end{lem}
\begin{proof}
Let us consider distinct $n\times k$-matrices of word with at most only one nonzero cell outside the last nonzero column $k$.

Let us begin from the matrices $V_{i,j}$ with unit in $(i,j)$ cell ($j<k$) and units in ($m,k$) cells for all $m$ except $i$.
The remaining cells contain zeros.
So we have $n-1$ units in the $k$-th column and only one unit in remaining $k-1$ columns of the matrix $V_{i,j}$.
Let the matrix $K$ have units in the $k$-th column and zeros in the other columns.
There are $n(k-1)$ matrices $V_{i,j}$. Together with $K$ they belong to the set $V$.
So we have $n(k-1)+1$ matrices. For instance,

\begin{picture}(0,40)
\end{picture}
$V_{1,1}=\left(
\begin{array}{cccccccc}
  1 & 0 & 0 & . & . & 0  \\
  0 & 0 & 0 & . & . & 1  \\
  0 & 0 & 0 & . & . & 1  \\
  . & . & . & . & . & .  \\
  0 & 0 & 0 & . & . & 1  \\
  0 & 0 & 0 & . & . & 1  \\
\end{array}
\right)$
\begin{picture}(4,40)
\end{picture}
$V_{3,2}=\left(
\begin{array}{cccccccc}
  0 & 0 & 0 & . & . & 1  \\
  0 & 0 & 0 & . & . & 1  \\
  0 & 1 & 0 & . & . & 0  \\
  . & . & . & . & . & .  \\
  0 & 0 & 0 & . & . & 1  \\
  0 & 0 & 0 & . & . & 1  \\
\end{array}
\right)$
\begin{picture}(4,40)
\end{picture}
$K=\left(
\begin{array}{cccccccc}
  0 & 0 & 0 & . & . & 1 \\
  0 & 0 & 0 & . & . & 1 \\
  0 & 0 & 0 & . & . & 1 \\
  . & . & . & . & . & . \\
  0 & 0 & 0 & . & . & 1 \\
  0 & 0 & 0 & . & . & 1 \\
\end{array}
\right)$

 The first step is to prove that the matrices $V_{i,j}$ and $K$ generate the space with the set $V$.
For arbitrary matrix $T$ of word from $V$ for every $t_{i,j} \neq 0$ and $j<k$,
let us consider the matrices $V_{i,j}$ with unit in the cell $(i,j)$ and the sum of them $\sum V_{i,j}=Z$.

The first $k-1$ columns of $T$ and $Z$ coincide.
   Hence in the first $k-1$ columns of the matrix $Z$ there is at most only one unit in any row.
 Therefore in the cell of $k$-th column of $Z$ one can find only value of $m$ or $m-1$.
The value of $m$ appears if there are only zeros
in other cells of the considered row. Therefore $\sum V_{i,j} - (m-1)K=T$.
Thus every matrix from the set $V$ is a span of $(k-1)n +1$ matrices from $V$.
It remains now to prove that the set of matrices $V_{i,j}$ and $K$ is a set of linear independent matrices.

If one excludes a certain matrix $V_{i,j}$ from the set of these matrices, then it is impossible
 to obtain a nonzero value in the cell $(i,j)$ and therefore to obtain the matrix $V_{i,j}$.
So the set of matrices $V_{i,j}$ is linear independent.
Every non-trivial linear combination of the matrices $V_{i,j}$ equal to a matrix of word has at
 least one nonzero element in the first $k-1$ columns.
Therefore, the matrix $K$ could not be obtained as a linear combination of the matrices $V_{i,j}$.
Consequently the set of matrices $V_{i,j}$ and $K$ forms a basis of the set $V$.
\end{proof}

\begin{cor}  \label {c2}
The set of all row monomial $n \times(n-1)$-matrices of words
(or $n\times n$-matrices with zeros in a fixed column)
has $(n-1)^2$ linear independent matrices.

The set of row monomial $n \times 2)$-matrices of words
has at most $n+1$ linear independent matrices.

The set of  row monomial matrices of words with one column
has at most $n$ linear independent matrices.

 \end{cor}

 \begin{cor}  \label {c3}
There are at most $n(n-1)+1$ linear independent matrices of words in the set of $n\times n$-matrices.
 \end{cor}

\begin{lem} \label{l3} { Distributivity}

For every words $b$ and $x_i$

\centerline{$M_b\sum\tau_iM_{x_i}=\sum\tau_i(M_bM_{x_i})$.}

\centerline{$(\sum\tau_iM_{x_i})M_b=\sum\tau_i(M_{x_i}M_b)$.}

\end{lem}

\begin{proof}
The matrix $M_b$ from left
shifts rows of every $M_{x_i}$ and of the sum of them in the same way according to Remark \ref{r4}.
$M_b$ removes common row of them and replace also by common row
(Remark \ref{r4}).

Therefore the matrices $M_bM_{x_i}$ and the sum
$\sum\tau_iM_bM_{x_i}$ has the origin rows with one unit from
$M_{x_i}$ and maybe in another order than
in its linear combination $\sum\tau_iM_{x_i}$,
The matrix $M_b$ from right shifts column $m$ of every $M_{x_i}$ and of the sum of them
in the same way (to the same column $k$) according to Remark \ref{r4}.
$M_b$ merges columns of the sum and of terms in the same way  too.  (Remark \ref{r4}).

Therefore the matrices $M_{x_i}M_b$ and the sum
$(\sum\tau_iM_{x_i})M_b$ has the origin columns
(sometimes merged) of $M_{x_i}$ and from its linear combination
$(\sum\tau_iM_{x_i})M_b$, with the same merged columns.

\end{proof}

\section{Linear independent matrices $M_u$}

\begin{lem} \label{v11}

Let the space $W$ be generated by linear independent
$n\times n-$matrices $M_u$ of words $u$
of restricted length $1\leq |u|\leq j$ and units only in first $k<n$ columns of $M_u$.

Then some matrix $M_{u\beta} \not\in W$ for generator $M_u$ of $W$
and some letter $\beta$, sometimes with $|R(v)|< |R(u)|$.
\end{lem}

 \begin{proof}

Assume the contrary: for every word $u$ with $1\leq |u|\leq j$  of generator
$M_{u_i}$ of $W$ and every letter $\beta$ with the length ($|u_i\beta|\leq  j+1$)
the matrix $M_{u_i\beta}$ in $W$.

For every matrix  $M_v \in W$

\centerline{$M_v=\sum \tau_i M_{u_i}$}
with generators $M_{u_i}$ in  $W$ such that
$|u_i|\leq j$ with units  in first $k<n$ columns.

Therefore by distributivity from Lemma \ref{l3}

\centerline{$M_vM_{\beta}=(\sum \tau_i M_{u_i})M_{\beta}=\sum \tau_i M_{u_i}M_{\beta}$}
for matrices $M_{u_i}M_{\beta}= M_{u_i\beta}$ with its units only in first  $k<n$  columns.

Therefore also $M_{v\beta}=M_vM_{\beta}=M_v \sum \tau_iM_{u_i\beta}$ belongs to $W$ and by
induction for every word $t$ the matrix $M_{u_i t}$ belongs to $W$ with its units only in first $k<n$ columns.

By induction, for every word $t$ the matrix $M_{vt} \in W$.

Contradiction for considered synchronizing automaton because for some word $t$ the matrix  $M_{vt}$
has nonzero column $n$.

\end{proof}

\begin{cor} \label{v12}

Let the sequence of spaces $W_j$ of dimension $j$ be ordered by inclusion by grow of $j$.
 The basis of $W_1$ contains one letter that maps a pair of states into one and
the basis of $W_j \supset  W_{j-1}$.
The space $W_j$ is extended by matrix $M_{u\beta}$ for a letter $\beta$ and some  $M_u$
from basis of $W_{j-1}$.

Then the length of the word $u$  of every generator  $M_u$ of $W_j$ is not greater than $j$.

Matrices of left subword of every generator of $W_j$ are linear independent because all
generators and its right subwords  were obtained by adding a letter from right.

\end{cor}

\section{The equation with unknown $L_x$}

\begin{df}
We denote

\centerline{$M_u \sim_q M_v$.}

if the columns of the state $\bf q$ of both matrices are equal.

If the set of cells with units in
the column $\bf q$ of the matrix $M_v$ is a subset of the
analogous set of the matrix $M_u$ then we write

\centerline{$M_v \sqsubseteq_q M_u$}

\end{df}

$A s = \bf q$ for synchronizing word $s$.

The solution $L_x$ of the equation
\begin{equation}
M_uL_x=M_s \label{ux}
\end{equation}
for synchronizing matrix $M_s$ and arbitrary $M_u$
must have units in the column of the state $\bf q$.

\begin{lem}  \label{l5}
Every equation $M_uL_x=M_s$ (\ref{ux}) has a solutions $L_x$ with at  least  $n\geq |R(u)|>0$ units in column $q$,
Every nonzero column $j$ of of $M_u$ corresponds a unit in the cell $j$ of column $q$.

For solution $L_x$ with only $|R(u)|$ units in column $q$ (a minimal solution) $L_x\sqsubseteq_q L_y$
for any other solution $L_y$ of (\ref{ux}).

There exists one-to-one correspondence between units in the column $q$ of minimal solution $L_x$
and the set $c_u$ of states.

\end{lem}

\begin{proof}
The matrix $M_s$ of rank one has nonzero column of the state $\bf q$.

For every nonzero column $j$ of $M_u$ with elements $u_{i,j}=1$ and $s_{i,q}=1$ in the matrix $M_s$ the cell $(j,q)$
must have unit in the matrix $L_x$.
So the unit in the column $q$ of matrix $M_s$ is a product of
every unit from the column $j$ of $M_u$ and unit in the sell $j$ of column $q$ of $L_x$.

The set $R(u)$ of nonzero columns of $M_u$ corresponds the set of cells of the column $q$ with unit of $L_x$.

Therefore  the minimal solution $L_x$ has in the column $q$ $|R(u)|$ units.

So to the column $q$ of every solution belong at least $|R(u)|$ units.
The remaining units of the solution $L_x$ belong to the next
columns, one unit in a row.
The remaining cells obtain zero.

Lastly every solution $L_x$ is a row monomial matrix of word.

Zeros in the column $q$ of minimal $L_x$ correspond zero columns of $M_u$.
Therefore for matrix $L_y$ such that $L_x \sqsubseteq_q L_y$
we have $M_uL_y=M_s$.
On the other hand, every solution $L_y$ must have units in cells of column $q$ that correspond nonzero columns of $M_u$.

Thus minimal $L_x$ has $|R(u)|$ unuts in column $q$ and the equality $M_uL_x=M_uL_y=M_s$ is
equivalent to $L_x \sqsubseteq_q L_y$.

The matrix $M_u$ has set $R(u)$ of nonzero columns and maps the automaton on the set $c_u$ of states
 and on the set of units in the column $q$ of minimal $L_x$.
%Units in the column $q$ of $L_y$ correspond some set of states $c_u$.

\end{proof}

Lemma \ref{l5} explains the following

\begin{rem} \label{r7}

Every permutation and shift of $m$ nonzero columns $M_u$
induces corresponding  permutation of the set of $m$ units in
the column $q$ of minimal solution $L_x$ of (\ref{ux}), and vice versa.

\end{rem}

\section{Allocation of linear independent matrices $L_x$}

\begin{lem} \label{m}

There exists allocation of matrix $M_s$ and $kn$ linear independent
solutions $L_x$ in first $k+1<n$ columns of $n\times n$-matrix $T$.

\end{lem}

\begin{proof}
The dimension of the space of matrices with $k+1$ nonzero columns is at most $kn+1$ cells
by Lemma \ref{v3}.

$M_s$ is placed at the beginning in column $\bf q$ because $As={\bf q}$.

 Cells without unit in $T$ let us call empty. First all cells in last $n-1$ columns are empty
after allocation of  $M_s$.

We continue extend set of linear independent $L_x$ using consistent allocation of its units.
Only one unit of $L_x$  will be placed in empty cell, the remaining units we place in
non-empty cells of allocation of former  $L_x$.
The goal is  a sequence of linear independent matrices $L_x$.
The unit in  empty cell guarantees linear independence matrix $L_x$ from previously allocated $L_x$.

There are at most $n$ linear independent matrices $M_u$ with $n-1$
units in the column $q$,
 last units of them belong to distinct rows outside $q$ ( Corollary \ref{c2}).
One can shift every such unit along row to column $2$ without
change linear independence.

 Then let us allocate linear independent matrices $L_x$ with  $|R(x)|=n-k$ units in
column $q$ and continue by growth of $k$ from two to $n-2$.

There are for every $k$ at most $nk+1$  linear independent matrices  $L_x$ in $k+1$ first columns
 $L_x$ together with $M_s$ by Lemma \ref{l3} and  $L_x$ with  $|R(x)|=n-k$ units in column $q$ is a
considered part of them.

So one has  at most $nk+1$  linear independent matrices in at most $nk+n$ cells of  first $k+1$ columns of $T$.
 ($M_s$ was allocated first in $n$ cells.)
The allocation  for fixed $k$ has three steps for the set of $L_x$ and its $n$ units.

At the beginning for given $k$ and considered $L_x$, $|R(u)|$ units of $L_x$ are allocated in cells of
column $q$ corresponding nonzero columns of matrix $M_u$ in equation $M_uL_x=M_s$.
So $L_x$ satisfies the equation (\ref{ux}) as  minimal solution.
Then we choose from remaining $n-|R(u)|$ units of $L_x$ only one unit for empty cell in possible minimal
column  $i \leq (k+1)$.
We reduce on such way the set of empty cells in  $k+1$ first columns and number of  $L_x$ of given $k$.
The remaining  units are allocated arbitrarily from column two in non-empty cells of previously allocated $L_x$.

Thus one can place at most $n(n-2)$ linear independent matrices $L_x$ together with $M_s$
 in at most $n((n-2)+1)$  cells of first $n-1$ columns with $|R(u)|$ units in column $q$ of every
$L_x$ corresponding $M_u$ from  (\ref{ux}).

\end{proof}

 \section{Theorems}

\begin{thm} \label{t}

The deterministic complete $n$-state synchronizing automaton
$A$ with strongly connected underlying graph over alphabet $\Sigma$ has synchronizing word in $\Sigma$
of length at most $(n-1)^2$.

The matrices of left subword of some synchronizing word of the automaton are linear independent.

\end{thm}

\begin{proof}
 We consider solutions $L_x$ of  (\ref{ux}) for linear independent words $u$ from Lemma \ref{v11}.
By Lemma \ref{m}, at most $kn$ linear independent matrices $L_x$  together with $M_s$
can be allocated in  first nonzero $k$ columns of some matrix, say $W$.

For $k=n-2$, one has a space of matrices $L_x$ and $M_s$ of dimension at most $n(n-2)+1$.
All matrices  $L_x$ with at least two units in column $q$ have allocation in first $n-1$ columns.
Therefore for remaining words $u$ there are only matrices $L_x$ with one unit in column $q$.

By Lemma \ref{m}, one unit of such $L_x$ of next $u$ belongs to column $q$,
another unit is  in last column $n$ and remaining units are in columns from $2$ to $n-1$.

We replace the matrix $M_u$ from the basis of $W$ by this $L_x$ with one unit in column $q$.
The dimension of new space is  $n(n-2)+1$ as before.

 By Lemma \ref{l5} corresponding word $u$ of such $L_x$ has $|R(u)|=1$, whence the matrix $M_u$
has one nonzero column of synchronizing word $u$ with $Au=\bf q$.
The length of $u$ is restricted by $n(n-2)+1$ by Corollary \ref{v12}.

Moreover, by Corollary \ref{v12}, the length of the word $u$  of every generator  of considered space is not
greater than $n(n-2)+1$ and all these generators have linear independent matrices of its left subword.
Hence the obtained synchronizing word $u$ has linear independent matrices of left subwords of $M_u$.
\end{proof}

\begin{cor} \label{c14}
For every integer $k<n$ of deterministic complete $n$-state synchronizing automaton $A$ with strongly
 connected underlying graph over alphabet $\Sigma$
there exists a word $v$ of length at most $n(k-1)+1$ such that $|Av|\leq n-k$.
\end{cor}

\begin{thm}\label{t2}
The deterministic complete $n$-state synchronizing automaton
$A$ with underlying graph
over alphabet $\Sigma$ has synchronizing word in $\Sigma$ of length at most $(n-1)^2$.
\end{thm}
Follows from Theorem \ref{t} because the restriction for strongly connected graphs can be omitted due to \cite{Ce}.

\begin{thm}\label{t4}
Suppose that $|\Gamma\alpha|<|\Gamma|-1$ for a letter $\alpha \in\Sigma$ in deterministic complete $n$-state
synchronizing automaton $A$ with underlying graph $\Gamma$ over alphabet $\Sigma$.

Then the minimal length of synchronizing word of the automaton is less than $(n-1)^2$.

\end{thm}

Proof.
We follow the proof of Theorem  \ref{t}.

The difference is that at the beginning of the proof the equation (\ref{ux}) has at least
 two linear independent nontrivial solutions for the matrix $M_{\alpha}$ of a letter $\alpha$
equal to the first word $u=\alpha$ of length one. Number of linear independent matrices $L_x$
is greater then corresponding $|u|$.  The difference remains on every step.

Hence we obtain finally synchronizing word of length less than $(n-1)^2$ with strongly
connected ideal $I$ .

Let us go to the case of not strongly connected underlying graph with $n-|I|>0$ states
outside  $I$.

This ideal has synchronizing word of length at most $(|I|-1)^2$ (Theorem \ref{t}).
There is a word $p$ of length at most $(n-|I|)(n-|I|+1)/2$ such that $Ap \subset I$.

$(|I|-1)^2 +(n-|I|)((n-|I|)+1)/2<(n-1)^2$.
Thus, the restriction for strongly connected automata can be omitted.

\section{Examples}
J. Kari \cite{Ka} discovered the following example of $n$-state automaton
with minimal synchronizing word of length $(n-1)^2$ for $n=6$.

\begin{picture}(130,70)
 \end{picture}
\begin{picture}(130,74)
\multiput(6,60)(64,0){2}{\circle{6}}
\multiput(6,13)(64,0){2}{\circle{6}}
 \multiput(22,56)(22,0){2}{a}
\multiput(16,19)(34,0){2}{a}
 \put(36,21){\circle{6}}
\put(36,48){\circle{6}}
 \put(7,14){\vector(4,1){28}}
\put(7,57){\vector(4,-1){26}}

\put(39,52){\vector(4,1){27}}
 \put(37,20){\vector(4,-1){28}}
\put(67,63){\vector(-1,0){57}}
 \put(36,64){a}
\put(67,12){\vector(-1,0){57}}
 \put(32,0){a}

\put(70,15){\vector(0,1){42}}
 \put(70,59){\vector(0,-1){42}}
\put(34,21){\vector(1,1){36}}
 \put(52,28){b}

  \put(76,22){b}
\put(76,10){3}
\put(76,60){0}
\put(27,25){5}
\put(43,38){2}

\put(25,37){b}
\put(36,48){\circle{10}}
\put(-6,10){4}
\put(0,20){b}
\put(-6,60){1}
\put(0,45){b}
\put(37,42){\vector(2,1){4}}

\put(6,60){\circle{10}}
\put(4,64){\vector(2,1){4}}

 \put(6,13){\circle{10}}
\put(7,7){\vector(2,1){4}}
 \end{picture}

The minimal synchronizing word

\centerline{$s=\it ba^2bababa^2b^2aba^2ba^2baba^2b$}
has the length at the \v{C}erny border.

Every line below presents a pair (word $u$, column $q$ of $L_x$) of linear independent
 $L_x$ from the sequence with evidently non-linear picture.

$(b, 111110)$ $R(u)=5$

$(ba, 111011)$

$(ba^2, 111101)$

$(ba^2b, 111100)$ $R(u)=4$

$(ba^2ba, 111010)$

$(ba^2bab, 011110)$

$(ba^2baba, 101111 )$ $R(v)=5$  (l01011 of $L_x$)

$(ba^2babab, 101110)$ $R(u)=4$

$(ba^2bababa, 110101)$

$(ba^2bababa^2, 011101)$

$(ba^2bababa^2b, 111000)$ $R(u)=3$

$(ba^2bababa^2b^2, 011100)$

$(ba^2bababa^2b^2a, 110111)$ $R(v)=5$  (101010 of $L_x$)

$(ba^2bababa^2b^2ab, 001110)$ $R(u)=3$

$(ba^2bababa^2b^2aba, 100011)$

$(ba^2bababa^2b^2aba^2, 011111)$  $R(v)=5$ (010101 of $L_x$)

$(ba^2bababa^2b^2aba^2b, 110000)$ $R(u)=2$

$(ba^2bababa^2b^2aba^2ba, 011000)$

$(ba^2bababa^2b^2aba^2ba^2, 101000)$

$(ba^2bababa^2b^2aba^2ba^2b, 001101)$ $R(v)=3$ (001100 of $L_u$)

$(ba^2bababa^2b^2aba^2ba^2ba, 100010)$ $R(u)=2$

$(ba^2bababa^2b^2aba^2ba^2bab, 000110)$

$(ba^2bababa^2b^2aba^2ba^2baba, 001011)$  $R(v)=3$ (000011 of $L_u$)

$(ba^2bababa^2b^2aba^2ba^2baba^2, 000101)$ $R(u)=2$

$(ba^2bababa^2b^2aba^2ba^2baba^2b=s, 100000)$  $R(s)=1$

By the bye, the matrices of left subwords of $s$ are simply linear independent.

\section*{Acknowledgments}
I would like to express my gratitude to Francois Gonze, Dominique Perrin, Marie B{\'e}al, Akihiro Munemasa,
Wit Forys, Benjamin Weiss, Mikhail Volkov and Mikhail Berlinkov for fruitful and essential remarks.

 \end{document}